\documentclass[12pt]{article}
\usepackage{epsfig}
\usepackage{array}
\usepackage{float}
\usepackage{amstext}
\usepackage{rotating}
\usepackage{a4}
\usepackage{a4wide}
\usepackage{cite}

\newcommand{\be}{\begin{eqnarray}}
\newcommand{\ee}{\end{eqnarray}}

\newcommand{\bfive}{$\bar 5_F$}
\def\nn{\nonumber}
\def\ltap{\ \raisebox{-.4ex}{\rlap{$\sim$}} \raisebox{.4ex}{$<$}\ }
\def\gtap{\ \raisebox{-.4ex}{\rlap{$\sim$}} \raisebox{.4ex}{$>$}\ }

\begin{document}

\title{%\vspace{-2cm}
%\hfill {\small \texttt{HRI-P-09-01-00x}} 
%\vskip 0.4cm
\Large 
Probing Seesaw in an Adjoint SUSY SU(5) Model at LHC}  
\author{
Ram Lal Awasthi\thanks{email: \tt ramlal@hri.res.in},~~~
Sandhya Choubey\thanks{email: \tt sandhya@hri.res.in},~~~
Manimala Mitra\thanks{email: \tt mmitra@hri.res.in}
\\\\
{\normalsize \it Harish--Chandra Research Institute,}\\
{\normalsize \it Chhatnag Road, Jhunsi, 211019 Allahabad, India }\\ \\ 
}
\date{ \today}
\maketitle
\vspace{-0.8cm}

\begin{abstract}
\noindent  
The SU(5) GUT model extended with fermions in the adjoint $24_F$ 
representation predicts triplet fermions in the 100 GeV mass range,
opening up the possibility of testing seesaw at LHC. However, once 
the model is supersymmerized, the triplet fermion mass is constrained 
to be close to the GUT scale for the gauge couplings to unify. We propose 
an extension of the SUSY SU(5) model where type II seesaw can be 
tested at LHC. In this model we add a matter chiral field in the 
adjoint $\hat{24}_F$ 
representation and Higgs chiral superfields in the symmetric 
$\hat{15}_H$ and $\hat{\bar{15}}_H$ representations. We call this 
the symmetric adjoint SUSY SU(5) model. The triplet scalar 
and triplet fermion masses in this model are predicted to be 
in the 100 GeV and $10^{13}$ GeV range respectively, while  
the mass of the singlet fermion remains unconstrained. 
This gives a type I plus type II plus type III seesaw 
mass term for the neutrinos.  The 
triplet scalars with masses $\sim 100$ GeV range can be produced at the LHC. 
We briefly discuss the collider phenomenology and 
predictions for proton decay in this model.

\end{abstract}

\newpage

\section{Introduction}

Despite its tremendous success, the Standard Model (SM) of 
particle physics with the SU(3)$_C\times$SU(2)$_L\times$U(1)$_Y$
gauge group, is generally regarded as the low energy 
limit of a more fundamental and complete theory. Among 
the major stumbling blocks of the SM is the 
obeservation of neutrino masses and mixing. The other major 
problem faced by the SM is the hierarchy problem. Since 
the electroweak scale is $10^{16}$ times smaller than the 
Planck scale, it is expected that the quantum corrections 
to the Higgs mass should be $\sim 10^{18}$, unless one 
plots to cancel quadratic radiative corrections by fine-tuning.

Grand Unified Theories (GUTs), based on higher gauge groups,  
allow for the unification of the quarks and leptons, and 
more importantly, the unification of the gauge couplings. 
Since the rank of SU(3)$_C\times$SU(2)$_L\times$U(1)$_Y$ is 
4, the smallest simple Lie Group which contains the SM gauge 
group as its subgroup is SU(5). 
The SU(5) GUT model was proposed long ago \cite{su5} 
with three copies of \bfive and $10_F$ representations 
containing the three generations of 
quarks and leptons, a $24_G$ containing the gauge bosons, 
and two Higgs representations  $5_H$ and 
$24_H$. The $24_H$ acquires a Vacuum Expectation Value (VEV) 
at a high scale, known as the GUT scale. This brings about 
spontaneous breaking of the SU(5) 
gauge group, thereby giving mass to 12 of the 24 gauge bosons 
contained in the $24_G$. These massive gauge bosons are known 
in the literature as the $X$ and $Y$ gauge bosons. The remaining 
12 massless gauge bosons belong to the SM. The $W$ and $Z$ 
subsequently get massive when SM is spontaneously broken to 
SU(3)$_C\times$U(1)$_{em}$ as a result of the VEV of the 
SM doublet Higgs contained in the $5_H$ of SU(5). 

While the minimal SU(5) GUT is simple and elegant, it fails 
on a few counts. Firstly, the SM gauge couplings $g_i$ ($i=1,2,3$) 
do not unify at one scale. While $g_2$ and $g_3$ unify at 
around $10^{17}$ GeV, the coupling $g_1$ unifies with $g_2$ 
much earlier. This problem can solved by introducing supersymmetry, 
which anyway is required in order to stabilize the Higgs mass 
and hence solve the gauge hierarchy problem. 
Another lacuna in the minimal SU(5) GUT concerns the generation 
of neutrino mass. The minimal SU(5) predicts the neutrinos to be 
massless. Neutrino masses can be generated by extending either the 
fermion sector or the Higgs sector. 
Presence of SU(5) singlet fermions can 
give rise to the type I seesaw mechanism \cite{type1}, while 
introduction of $15_H$ can produce neutrino mass via the 
type II seesaw \cite{type2}. A lot of recent interest has been generated 
from the possibility of a third kind of seesaw in SU(5). This 
so-called type III seesaw \cite{type3}
can be realized by extending SU(5) 
with fermions in the adjoint representation, $24_F$ 
\cite{ma,adjointsu5,dp,readjsu5, adjsu5lepto}. 
This model has been popularly called the {\it adjoint SU(5)}. 
The $24_F$ contains the (1,3,0) fermion representation 
of the SM, which can mediate 
type III seesaw. 
The additional advantage one gets by introducing 
the $24_F$ is that the contribution of the representations of 
$24_F$ to the individual SM gauge coupling running is such that 
the gauge couplings unify at around $10^{15.5}$ GeV, {\it even without 
invoking supersymmetry}. This happens 
when the (1,3,0) fermion 
representation has mass in the 100s of GeV range, making it accessible 
at LHC \cite{type3lhc, aguila}. 
This opens up the possibility of testing seesaw at LHC.
A further intrinsic problem in minimal SU(5) concerns 
the masses of the d-type quarks and charged leptons, which become 
degenerate at the GUT scale. This is a well known problem and can be 
easily solved either by allowing higher dimensional operators 
or an additional $45_H$ to break this unwanted degeneracy. 

While requirement of supersymmetry for achieving gauge coupling  
unification in SU(5) is alleviated by adding a $24_F$, supersymmetry 
is still required for addressing the issue of the hierarchy 
problem. Therefore, one should consider a supersymmetric GUT 
as the complete theory. The supersymmetric version of the 
adjoint SU(5) has been proposed in the literature 
and has been called adjoint SUSY SU(5) 
\cite{susyadjsu5}. 
In this paper we study the issue of gauge coupling 
unification and the condition it imposes 
on the particle spectra. We find that once supersymmetry is 
imposed, the mass of the (1,3,0) fermion representation of 
$24_F$ turns out to be very close to the GUT scale,  
making it impossible 
to produce it at the LHC. Therefore, type III seesaw predicted 
by the adjoint SUSY SU(5) model cannot be 
tested at the current and even future collider experiments.

We next embark upon constructing a model based on the 
supersymmetric SU(5) which allows for TeV-scale seesaw mechanism 
that can be probed at the LHC. As mentioned above, addition 
of symmetric 
$15_H$ representation to minimal SU(5) 
allows for the type II seesaw mechanism. 
The ramifications of the $15_H$ representation for the 
gauge coupling unification in SU(5) without supersymmetry 
has been studied \cite{symmsu5_0,symmsu5,symmsu5_1}. In the absence of 
supersymmetry, gauge coupling unification can be 
achieved if the (3,2,1/6) representation of $15_H$ 
has mass in the $10^2$-$10^3$ GeV range, making them accessible at 
the LHC. However, the (1,3,1) scalars which mediate 
type II seesaw have masses in the intermediate range 
and are therefore inaccessible at the LHC. In addition, once 
supersymmetry is imposed, even the masses of the leptoquarks 
should be around the GUT scale in order to get gauge coupling 
unification. 

In this paper we propose a 
supersymmetric SU(5) GUT model where we add 
a matter chiral field in the 
adjoint $\hat{24}_F$ 
representation and Higgs chiral superfields in the symmetric 
$\hat{15}_H$ and $\hat{\bar{15}}_H$ representations. 
We call this ``symmetric adjoint SUSY SU(5)'' model. 
Since this model has (1,1,0) and (1,3,0) fermionic representations 
as well as (1,3,1) scalar multiplet, neutrino masses 
could get contributions from 
type I, type II, as well as type III seesaw. We show that 
gauge coupling unification constrains the particle masses 
such that the triplet scalar and the triplet fermion masses in this model are predicted to be 
in the 100 GeV and $10^{13}$ GeV range, respectively. The 
triplet scalars with masses $\sim 100$ GeV range can be produced at the LHC 
making seesaw testable at the LHC. We briefly discuss the phenomenological 
aspects of this model. 

The paper is organized as follows. In section 2 we briefly outline the particle 
content of the SUSY SU(5) GUT model and some of its extensions that 
are relevant to this paper. In section 3 we study the RG evolution of the 
SM gauge couplings and the constraints it imposes on the particle 
masses in the adjoint SU(5) and the adjoint SUSY SU(5). In section 4 
we propose the symmetric adjoint SUSY SU(5). We give the particle mass 
spectra expected in this model and show that these particle masses consistently 
give gauge coupling unification at the GUT scale. In section 5 we 
study the phenomenological consequences of this model. Finally, in 
section 6 we end with our conclusions.

\section{SUSY SU(5) and its Extensions}

We begin with a very brief overview of the particle content of 
the supersymmetric SU(5) model. The minimal version of the 
model \cite{susysu5} is comprised of three families of 
$\hat{\bar{5}}\equiv (\bar{3},1,1/3)\oplus (1,2,-1/2) 
\equiv (\hat{d}^C, \hat{L})$ and 
$\hat{10}\equiv (\bar{3},1,-2/3)\oplus (3,2,1/6)\oplus (1,1,1) 
\equiv (\hat{u}^C, \hat{Q}, \hat{e}^C)$ matter chiral 
multiplets, while the Higgs sector comprises of a 
$\hat{5}_H \equiv (3,1,-1/3) \oplus (1,2,1/2)\equiv (\hat{T}, \hat{H_u})$, a 
$\hat{\bar{5}}_H \equiv (\bar{3},1,1/3)\oplus (1,2,-1/2)
\equiv (\hat{\bar{T}}, \hat{{{H}_d}})$ and 
a $24_H \equiv (8,1,0)\oplus (1,3,0)\oplus (3,2,-5/6)
\oplus (\bar{3},2,5/6)\oplus (1,1,0)
\equiv (\hat{\Sigma}_8,\hat{\Sigma}_3,\hat{\Sigma}_{(3,2)},
\hat{\Sigma}_{(\bar{3},2)},\hat{\Sigma}_{0})$. The gauge sector 
is obviously contained in the $24_G$, which has 12 gauge bosons 
belonging to the SM and another 12 which are new (called the $X$ and 
$Y$ gauge bosons) and get mass 
at the SU(5) breaking scale. 
In this minimal 
model, SU(5) is broken when the $\Sigma_{0}$ Higgs picks up a 
VEV at around $10^{15.5}$ GeV. 
%It is easy to see that 
%the $X$ and $Y$ gauge bosons would lead to dimension 6 operators 
%at the electroweak scale which break both baryon and lepton numbers. 
%This would therefore trigger proton decay. 
%However, all non-SM particles including the $X$ and $Y$ gauge bosons 
%have masses at or close to the GUT scale, thereby evading the 
%proton decay bounds which is currently at $1.01\times 10^{34}$ years 
%\cite{pdk}. Presence of supersymmetry allows the creation of 
%dimension 5 effective operators which mediate proton decay \cite{susypdk}. 
%However, it has been shown that a large region of parameter space exists 
%which still allows for the effect of these operators to be within 
%control, making the model experimentally viable. 

As discussed in the introduction, the minimal SUSY SU(5) is still 
incomplete, as it does not give neutrino mass. In order to generate 
neutrino masses one needs to extend the model with additional 
multiplets. In the adjoint SUSY SU(5) model, one introduces additional  
matter chiral fields in the adjoint representation 
$\hat{24}_F \equiv (8,1,0)\oplus (1,3,0)\oplus (3,2,-5/6)
\oplus (\bar{3},2,5/6)\oplus (1,1,0)
\equiv (\hat{\rho}_8,\hat{\rho}_3,\hat{\rho}_{(3,2)},
\hat{\rho}_{(\bar{3},2)},\hat{\rho}_{0})$. Neutrino masses are 
generated via the type I (mediated by $\rho_{0}$) and 
type III (mediated by $\rho_3$) seesaw. Alternatively, one could also  
introduce the $\hat{15}_H \equiv (1,3,1)\oplus (3,2,1/6) \oplus (6,1,-2/3) 
\equiv (\hat{\Delta}_3, \hat{\Delta}_{(3,2)}, \hat{\Delta}_{6})  $ 
Higgs chiral multiplet. The $\Delta_3$ could give rise to neutrino 
masses by the type II seesaw mechanism.

\section{RG running of gauge couplings in Adjoint SUSY SU(5)}
\label{sec:rgsu5}

The RG evolution of the three SM gauge couplings between the 
electroweak scale and the GUT scale 
in the $\overline{MS}$ scheme 
up to one loop corrections 
and including threshold effects is given by 
\be
 2\pi (\alpha_{i}^{-1}(M_Z)-\alpha_{G}^{-1}(M_G))= b_i \ln(\frac{M_G}{M_Z})
   +\sum_j b_i(j)\ln(\frac{M_G}{M_j})
\label{eq:rge}
\ee 
where the second term in the RHS of Eq. (\ref{eq:rge}) 
give threshold corrections at one loop coming from particles whose 
mass $M_j > M_Z$. 
The above 
equation relates the gauge couplings at $M_Z$ to the 
gauge coupling at the GUT scale $M_G$. 
The 
$b_i$'s are the co-efficients of the $\beta$-functions 
for gauge couplings $g_i$ at the one loop level. 
They can be calculated for a 
given gauge group ${\cal G}_i\otimes {\cal G}_j$ as
\be
b_i&=&-\frac{11}{3}T_G(R_i)d(R_j)+
\frac{2}{3}T_F(R_i)d(R_j)+\frac{1}{3}T_S(R_i)
d(R_j)
~,
%\nonumber \\
%&=&b_i^G + b_i^F + b_i^S
\label{eq:bi}
\ee
where $T(R_i)$ are 
the Casimir of the representation $R_i$ under the 
group ${\cal G}_i$ and $d(R_j)$ is the 
dimension of the representation $R_j$ under group ${\cal G}_j$. The 
first, second and third terms in Eq. (\ref{eq:bi}) 
gives the contribution coming 
from gauge bosons, fermions and scalars in the model. 

\begin{table}
\begin{center}
\begin{tabular}{|l|c|c|c|r|}
\hline
SM multiplet & Particle type & $b_{U(1)_Y}$ & $b_{SU(2)_L}$ & $b_{SU(3)_C}$\\

\hline
(1,1,0)   &  All   &0& 0      &0\\
\hline
         & G&-11/5& 0& 0\\ \cline{2-5}
(1,1,1)  & F&2/5 &0&0\\ \cline{2-5}
         & S&1/5 &0 &0 \\ \cline{2-5}
\hline

             &  G   &-11/10&-11/6&0\\ \cline{2-5}
(1,2,1/2) \& (1,2,-1/2)   &  F   &1/5& 1/3      &0 \\ \cline{2-5}
             &  S   &1/10& 1/6& 0\\ \cline{2-5}

\hline 

            &  G  &0&-22/3&0\\ \cline{2-5}
(1,3,0)     &  F  &0 &4/3& 0 \\ \cline{2-5}
             &  S & 0  &1/3& 0 \\ \cline{2-5}            

\hline
            &  G   &0&0&-11\\ \cline{2-5}
(8,1,0)   &  F  &0 &0&2 \\ \cline{2-5}
             &  S & 0  &0&1/2 \\  \cline{2-5}

\hline

             &  G   &-11/15&0&-11/6\\ \cline{2-5}
(3,1,-1/3) \& ($\overline{3}$,1,1/3)   &  F   &2/15& 0      &1/3\\ \cline{2-5}
             &  S   &1/15& 0      &1/6\\ \cline{2-5}
\hline    
             &  G  & -11/30   & -11/2  & -11/3    \\ \cline{2-5}
(3,2,1/6) \& ($\overline{3}$,2,-1/6)    &  F  &  1/15  & 1  &  2/3   
\\ \cline{2-5}
             &  S  & 1/30   & 1/2  & 1/3     \\ \cline{2-5} 

\hline

             &  G   &-44/15&0&-11/6\\ \cline{2-5}
($\overline{3}$,1,-2/3) \& (3,1,2/3)   &  F   &8/15& 0      &1/3\\ \cline{2-5}
             &  S   &4/15& 0      &1/6\\ \cline{2-5}
\hline    
         
            &G&-55/6&-11/2&-11/3\\ \cline{2-5}
(3,2,-5/6) \& ($\overline{3}$,2,5/6)&F&5/3&1&2/3\\ \cline{2-5}
                     &S&5/6&1/2&1/3\\ \cline{2-5}

\hline   
\hline

        & G & -33/5   &  -22/3   & 0   \\ \cline{2-5}
 (1,3,1) \& (1,$\overline{3}$,-1)& F &  6/5   & 4/3   & 0  \\ \cline{2-5}
        & S  & 3/5   &  1/3   &  0     \\ \cline{2-5}

\hline
          & G & -88/15  & 0  & -11/2   \\ \cline{2-5}
(6,1,-2/3) \& ($\overline{6}$,1,2/3) &F& 16/15  & 0 & 1  \\ \cline{2-5}
          & S & 8/15   & 0  & 1/2 \\ \cline{2-5}
\hline
\end{tabular}
\end{center}
\caption{\label{tab:bivalues}
The $\beta$-function co-efficients $b_i$  for the different SM representations. 
The SM representation is given in first column. The `G', `F' and `S' in the second 
column stand for gauge bosons, fermions and scalars, respectively. 
}
\end{table}
 
In Table \ref{tab:bivalues} we give the individual contributions 
to $b_i$ for the different possible SM representations. We give 
these values separately for gauge bosons (G), fermions (F) and 
scalars (S) for the three gauge couplings. One can note from 
the Table \ref{tab:bivalues} that the contribution $b_i$ for 
gauge bosons is always negative, while that for fermions and 
scalars is always positive. Since any extension of a given 
gauge theory entails addition of matter and Higgs fields without 
tampering with the gauge sector, and since every fermion and 
Higgs field brings a further negative contribution on the 
RHS of Eq. (\ref{eq:rge}), extension of any GUT model 
results in reducing the slope of $\alpha_i^{-1}$ as a 
function of the energy scale. The extent of this reduction 
for a given gauge coupling $g_i$ 
depends on the representation of the relevant multiplet 
under the gauge group ${\cal G}_i$. Therefore, even though 
every additional multiplet has the effect of reducing the 
slope of all the three $\alpha_i^{-1}$, the relative 
reduction between them is different for different 
multiplets. 

If one imposes unification and eliminates the unified gauge 
coupling $\alpha_{G}^{-1}(M_G)$ from the Eq. (\ref{eq:rge}) 
for the three SM gauge couplings, 
one gets two equations 
%\cite{hall91}
\be
\ln\frac{M_G}{M_Z} = \frac{\Delta_{12}}{{\cal B}_{12}}\,,
\label{eq:gutscale}
\ee
\be
{\cal B}_{23} = \frac{\Delta_{23}}{\Delta_{12}}\,{\cal B}_{12}\,,
\label{eq:b23b12gen}
\ee
with, 
\be
{\cal B}_{ij} = B_{ij} + \sum_{k} B_{ij}(k)\,\frac{\ln({M_G}/{M_k})}
{\ln({M_G}/{M_Z})}\,,
\label{eq:Bij}
\ee
where, $B_{ij} = b_i - b_j$ and 
$\Delta_{ij} = 2\pi(\alpha_i^{-1} -\alpha_j^{-1})$. Inserting the 
experimental measured values of $\alpha_i^{-1}$ at $M_Z$
($\alpha_1^{-1}(M_Z)=58.85$, 
$\alpha_2^{-1}(M_Z)=29.46$, $\alpha_3^{-1}(M_Z)=8.50$), 
we get
\be
{\cal B}_{23} = 0.713\,{\cal B}_{12}\,.
\label{eq:b23b12expt}
\ee
One can use Table \ref{tab:bivalues} to calculate the $b_i$'s for the 
SM and the MSSM. These values come out to be  
\be
b_i^{SM}\equiv\left(\frac{41}{10},-\frac{19}{6},-7\right)\,, 
\label{eq:smb}
\ee
\be
b^{MSSM}_i=
\left(\frac{33}{5},1,-3\right)
~.
\label{eq:mssmb}
\ee
Using Eq. (\ref{eq:smb}) one gets ${\cal B}_{23} / {\cal B}_{12} = 0.53$ 
for the SM, which is inconsistent with Eq. (\ref{eq:b23b12expt}). 
Therefore, unification fails in the SM. For the MSSM on the 
other hand one gets using Eq. (\ref{eq:mssmb}) 
${\cal B}_{23} / {\cal B}_{12} = 0.714$ nearly 
consistent
%\footnote{The values of ${\cal B}_{23}$ and ${\cal B}_{12}$ 
%used above are assuming that all particles are massless. However, 
%below the supersymmetry scale, the superparticles will acquire 
%masses in the TeV range giving small threshold corrections which 
%help reducing the value of ${\cal B}_{23} / {\cal B}_{12}$ and making 
%it closer to the experimental value.} 
with Eq. (\ref{eq:b23b12expt})
and hence the experimental values of the coupling constants at the 
electroweak scale. 

For the SM all particles are massless above the electroweak scale 
and hence ${\cal B}_{ij} = b_i - b_j$. As discussed above 
this fails to give unification of the gauge couplings. 
However, if one extends the model to include other massive 
particles which contribute to ${\cal B}_{ij}$ such that 
the increase $\Delta({\cal B}_{23})$ is greater than the 
increase $\Delta({\cal B}_{12})$, then one could get 
unification even without supersymmetry.  
In the minimal SU(5) all the massive particles above the 
electroweak scale are at the GUT scale and hence do not contribute 
to the gauge coupling running. In order to achieve unification 
one must therefore add additional matter multiplets with masses 
in the intermediate scale such that 
${\cal B}_{23} / {\cal B}_{12} $ can be raised. 
One could add either a $15_H$ to the minimal SU(5) or 
a $24_F$ in order to achieve unification. While the former 
leads to type II seesaw, the latter gives rise to type I+III seesaw.  
In the rest of this section we will focus on the SU(5) 
extensions with $24_F$ as it allows for the testing of 
the seesaw framework at the LHC. We remind the reader than in 
models with $15_H$ \cite{symmsu5_0,symmsu5,symmsu5_1}
one gets the 
triplet scalar $\Delta_3$ at the intermediate scale and 
hence unobservable at colliders. 

The non-supersymmetric adjoint SU(5) with one family of 
$24_F$ receives threshold correstions from $\rho_8\equiv 
(8,1,0)$, $\rho_3\equiv (1,3,0)$ and $\rho_{(3,2)} \equiv (3,2,-5/6)$. 
The $\rho_8$ contributes to the running of $\alpha_3^{-1}$ only while 
$\rho_3$ impacts the running of $\alpha_2^{-1}$ only. The effect 
of $\rho_{(3,2)}$ on the other hand is felt by all the three. If 
one restricts $\rho_{(3,2)}$ to have mass close to the GUT scale 
then its impact on the RG running of gauge couplings can be 
reduced. Therefore, only $\rho_3$ gives threshold corrections 
to the running of $\alpha_2^{-1}$ and since it is a fermion 
its $b_2$ contribution helps to reduce the value of $ {\cal B}_{12} $.  
Likewise, the effect of threshold corrections due to $\rho_8$ 
impacts the running of $\alpha_3^{-1}$ and together $\rho_3$ 
and $\rho_8$ can be constrained to have masses below the 
GUT scale such that ${\cal B}_{23} / {\cal B}_{12} = 0.713$ 
and unification constraint can be satisfied. In this model $ {\cal B}_{12} $ gets 
threshold contribution from  $\rho_3$ and $\rho_{3,2}$ and  one can easily 
relate the GUT scale $M_G$ to $M_{\rho_3}$. Using  
Eq. (\ref{eq:gutscale}) and Table \ref{tab:bivalues} one gets 
\be
\log M_G = 16.4 - 0.3\log M_{\rho_3}
\,.
\ee
For $M_G = 10^{15.8}$ GeV the above equation gives 
$M_{\rho_3} = 10^2$ GeV.  The Eq. (\ref{eq:b23b12gen}) can 
next be used to find the mass of $\rho_8$ as 
\be
\ln M_{\rho_8} = 5.7 + \log M_{\rho_3}
\,,
\ee
leading to the relation $M_{\rho_8}/M_{\rho_3} \sim 10^{5.7}$ 
between the masses of the 
two particles. Therefore, while 
production of $\rho_8$ is impossible at LHC, the production of $\rho_3$ 
is possible via gauge interactions, making it possible to 
probe seesaw.

We next turn our attention at the SUSY adjoint SU(5) and impose 
the unification constraints given by Eqs. (\ref{eq:gutscale}) 
and (\ref{eq:b23b12gen}). In the SUSY adjoint SU(5) 
we have an extra $\hat{\bar 5}_H$ multiplet. In addition, 
for every particle contribution we have to include the 
contributions coming from the corresponding superparticle. 
We have already seen that in the MSSM itself these 
additional superparticle contributions are enough to 
get gauge coupling unification at the right scale. 
In SUSY adjoint SU(5) we get threshold corrections from the 
additional particles as well. Assuming all superpartners of 
the standard model particles to be at the TeV scale, 
all multiplets of $\hat{24}_H$, leptoquark 
of $\hat{24}_F$, and Higgs triplet $T$ of $\hat{5}_H/\hat{\bar{5}}_H$ 
at the GUT scale, 
one gets using 
Eq. (\ref{eq:gutscale})
\be
\log M_G = 22.8 - 0.42\log M_{\rho_3}
\,.
\ee
As a result, for $M_G\sim 10^{16}$ GeV we get $M_{\rho_3} \sim 10^{16}$ GeV. 
Therefore, testing seesaw at LHC will be impossible once SUSY is 
imposed in the adjoint SU(5) model. 

\begin{figure}[t]
\begin{center}
\includegraphics[width=0.5\textwidth,angle=270]{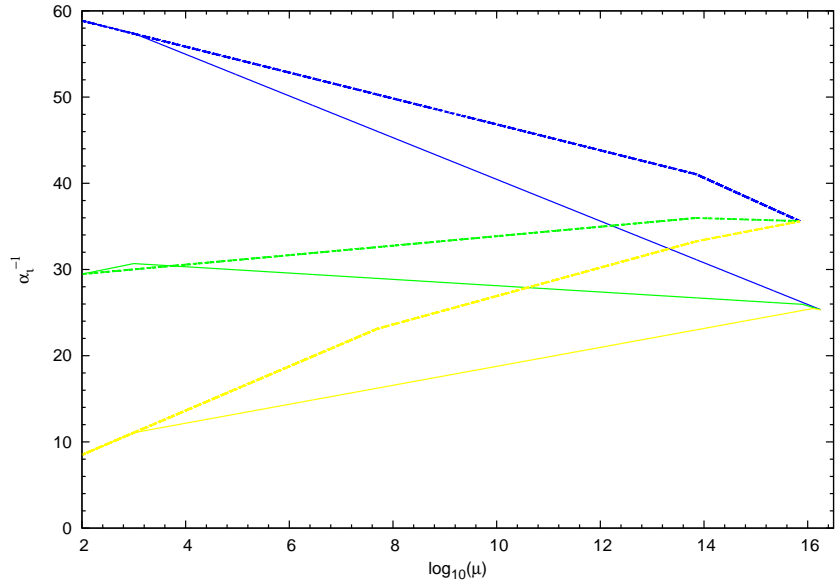}
\end{center}
\caption{\label{fig:rgsu5}
Gauge coupling running for the adjoint SU(5) (dashed lines) 
and adjoint SUSY SU(5) (solid lines) models. 
For the adjoint SU(5) we have taken 
$M_{\rho_3}=100$ GeV, $M_{\rho_8}=10^{7.7}$ GeV  and 
$M_{\rho_{(3,2)}}=5\times 10^{14}$, while all other parameters are at the GUT scale. 
For the adjoint SUSY SU(5) model $M_{SUSY}= 1$ TeV while all other particles 
have masses close to the GUT scale.  }
\end{figure}

We show in Fig. \ref{fig:rgsu5} the running of the gauge 
couplings for the adjoint SU(5) (thick lines) 
and the adjoint  SUSY SU(5) (thin lines)
models. The kinks in the running show the position of the 
masses of the SU(2) triplet and the color octet fermions. 
Note that the unified gauge coupling $\alpha_G^{-1}$ is 
lower for the adjoint SUSY SU(5). This is in fact a generic 
feature of any GUT model which is extended by adding more 
multiplets. Extending the multiplet content of the model 
always results in additional fermion and scalar contributions, 
both of which lower the slope of $\alpha_i^{-1}$ as well 
as the value of $\alpha_G^{-1}$.

%%%%%%%%%%%%%%%%%%%%%%%%%%%%%%%%
%%%%%%%%%%%%%%%%%%%%%%%%

\section{The Symmetric SUSY Adjoint SU(5)}

In this section we propose a supersymmetric 
SU(5) GUT model which generates 
neutrino masses by the seesaw mechanism, such that the 
mass of the 
seesaw mediating particle(s) is in the TeV range, making 
it possible to produce them at the LHC, and hence 
test the seesaw mechanism. 
We propose to extend the 
SUSY SU(5) by adding three additional multiplets, a 
matter chiral supermultiplet $\hat{24}_F$ and the  
Higgs chiral  supermultiplets $\hat{15}_H$ and $\hat{\bar{15}}_H$. Since 
$15$ is symmetric and $24$ the adjoint representation
under SU(5), we name this model the 
{\it symmetric adjoint SUSY SU(5)}. 
The renormalizable superpotential for this model is given by
\be
{\cal W}_{ren}&=& Y \hat{\overline{5}}_F\hat{10}_F\hat{\overline{5}}_H+ 
Y'\hat{10}_F\hat{10}_F\hat{5}_H+
Y_{\Delta}\hat{\overline{5}}_F\hat{15}_H \hat{\overline{5}}^T_F
+Y_{\rho}\hat{\overline{5}}_F\hat{24}_F\hat{5}_H+m_{H}\hat{\overline{5}}_H\hat{
5}_H+\mu_{\Delta}\hat{\overline{5}}_H\hat{15}_H\hat{\overline{5}}_H^T 
\nn \\
&+&\mu_{\bar{\Delta}}\hat{5}_H^T\hat{\overline{15}}_H\hat{5}_H+m_{\rho}
Tr(\hat{24}_F^2)+\lambda_{\rho}Tr(\hat{24}_F^2\hat{24}_H)
+m_{\Sigma}Tr(\hat{24}_H^2)+\lambda_{\Sigma} Tr(\hat{24}_H^3)\nn
\\
&+&Y_{\Sigma}\hat{\overline{5}}_H\hat{24}_H\hat{5}_H+m_{\Delta}
Tr(\hat{\overline{15}}_H\hat{15}_H)+\lambda_{\Delta}'Tr(\hat{15}_H\hat{
\overline{15}}_H\hat{24}_H)+\lambda_{\Delta}''Tr(\hat{15}_H\hat{24}^T_H
\hat{\overline{15}}_H)\,\nn\\
&+&Y^{\prime\prime} \hat{24}_F \hat{10}_F \hat{\overline{15}}_H\,.
\label{eq:spot}
\ee
The SU(5) gauge symmetry breaks to the SM 
when the $24_H$ Higgs gets a VEV. Subsequently, the VEV of 
$H_u$ and $H_d$ in ${5}_H$ and $\bar{5}_H$ 
breaks SM and generates the seesaw type I and 
type III masses for the neutrinos. In addition, the VEV of 
$\Delta_3$ in $15_H$ generates the type II seesaw masses 
for the neutrinos. Therefore, in our model neutrinos could get 
masses from all the three types of seesaw mechanisms. 
In what follows, we will first extract the mass spectrum of 
the particles in this model. We next look at the RG 
running of the SM gauge couplings

\subsection{Particle Mass Spectra} 
%of $\hat{24}_F$ and $\hat{15}_H(\hat{
%\overline{15}}_H)$}

%\subsubsection{$\hat{24}_F$ Mass Spectrum}

The relevant super-potential to generate mass of $\hat{24}_F$ particles 
is
\be
{\cal W}_{\rho}^{M}&=&m_{\rho}Tr(\hat{24}_F^2)+\lambda_{\rho}
Tr(\hat{24}_F^2\hat{
24}_H) + \frac{\gamma_{F}}{\Lambda}Tr(\hat{24}_F^2 \hat{24}_H^2) + 
\frac{\delta_{F}}{\Lambda}Tr(\hat{24}_F^2)Tr( \hat{24}_H^2) +
\nonumber \\ &&
\frac{\lambda_{F}}{\Lambda}Tr(\hat{24}_F \hat{24}_H\hat{24}_F \hat{24}_H) +
\frac{\lambda'_{F}}{\Lambda}[Tr(\hat{24}_F \hat{24}_H)]^2
\,,
\label{eq:wrhom}
\ee
where we have also included the dimension five effective operators that contribute to the 
mass of the particles belonging to $\hat{24}_F$. The scale $\Lambda$ could be associated 
with the Planck scale. 
The higher dimension effective operators have to 
be added anyway in the SU(5) GUT in order to break the degeneracy between the masses 
of the charged lepton and the d-type quark masses. The above leads to 
the following expressions for the 
masses of $\hat{\rho}_{0}$, $\hat{\rho}_3$, $\hat{\rho}_8$ and $\hat{\rho}_{(3,2)} $
\be
M_{\rho_{0}}&=&m_{\rho}-\frac{\lambda_{\rho}v_{\Sigma}}{\sqrt{30}} +
\frac{v_\Sigma^2}{\Lambda}\bigg(\delta_{F} + \lambda'_{F} + \frac{7}{30}(\gamma_{F} + \lambda_{F})\bigg )
\,,\nn \\
M_{\rho_3}&=&m_{\rho}-\frac{3\lambda_{\rho}v_{\Sigma}}{\sqrt{30}}+
\frac{v_\Sigma^2}{\Lambda}\bigg(\delta_{F} + \frac{3}{10}(\gamma_{F} + \lambda_{F})\bigg )\,,\nn \\
M_{\rho_8}&=&m_{\rho}+\frac{2\lambda_{\rho}v_{\Sigma}}{\sqrt{30}}+
\frac{v_\Sigma^2}{\Lambda}\bigg(\delta_{F} + \frac{2}{15}(\gamma_{F} + \lambda_{F})\bigg )\,, \nn \\
M_{\rho_{(3,2)}}&=&m_{\rho}-\frac{\lambda_{\rho}v_{\Sigma}}{2\sqrt{30}}+
\frac{v_\Sigma^2}{\Lambda}\bigg(\delta_{F} + \frac{(13 \gamma_{F} - 12 \lambda_{F})}{60}\bigg )
\,, 
\label{eq:rhomass}
\ee
respectively. The higher dimensional terms bring additional parameters such that we 
can tune all the above particles to be at any arbitrary scale. We will see below that to be 
able to test seesaw at LHC one should allow for $\hat{\rho}_3$ to have masses $\sim 10^{13}$ GeV and 
$\hat{\rho}_8$ at some intermediate scale, while the mass of $\hat{\rho}_{(3,2)}$ 
is constrained to be at the GUT scale.

The corresponding terms in the superpotential which contribute to the 
mass of the $\hat{24}_H$ are given by
\be
{\cal W}_\Sigma^M = \alpha\,Tr(\hat{24}_H^2) +\beta\,Tr(\hat{24}_H^3)+\frac{\gamma_H}{\Lambda}Tr(\hat{24}_H^4)
+\frac{\delta_H}{\Lambda}(Tr(\hat{24}_H^2))^2
\,,
\label{eq:sp24H}
\ee
where in the third and the fourth terms we have introduced higher dimensional effective operators 
suppressed by some heavy scale $\Lambda$. The masses of the 
$\hat{\Sigma}_8$, $\hat{\Sigma}_3$ and $\hat{\Sigma}_0$ then turn out to be
\be
M_{\Sigma_8} &=& \alpha + \frac{6\beta v_\Sigma}{\sqrt{30}} + \frac{4\gamma_H v_\Sigma^2}{5\Lambda}
+ \frac{2\delta_H v_\Sigma^2}{\Lambda}
\,,\nn\\
M_{\Sigma_3} &=& \alpha - \frac{9\beta v_\Sigma}{\sqrt{30}} + \frac{9\gamma_H v_\Sigma^2}{5\Lambda}
+ \frac{2\delta_H v_\Sigma^2}{\Lambda}
\,,\nn\\
M_{\Sigma_0} &=& \alpha - \frac{3\beta v_\Sigma}{\sqrt{30}} + \frac{7\gamma_H v_\Sigma^2}{5\Lambda}
+ \frac{6\delta_H v_\Sigma^2}{\Lambda}
\,.
\ee
With four free parameters, $\alpha$, $\beta$, $\gamma$ and $\delta$, we can tune 
the masses of the particles such that $\hat{\Sigma}_3$ and $\hat{\Sigma}_0$ are at the 
GUT scale while $\hat{\Sigma}_8$ is light, and at the same time satisfy the minimization 
condition of the scalar potential required for the spontaneous breaking of the SU(5) 
gauge symmetry. 
%We have checked that the additional higher order terms in 
%Eq. (\ref{eq:sp24H}) does not significantly alter the minimization condition of 
%$24_H$. 

Finally, we give the mass spectrum of $\hat{15}_H$. 
Following terms from the superpotential Eq. (\ref{eq:spot}) 
give contribution to the masses of the $\hat{15}_H$ (and 
$\hat{\bar{15}}_H$) multiplets:
\be 
{\cal W}_{\Delta}^{M}=m_{\Delta}Tr(\hat{15}_H\hat{\overline{15}}_H)
+\lambda_{\Delta}'Tr(\hat{15}_H\hat{\overline{15}}_H\hat{24}_H)+\lambda_{\Delta}''
Tr(\hat{15}_H\hat{24}_H^T\hat{\overline{15}}_H)\,.
\label{eq:higgstrip}
\ee
We have not shown the higher dimensional operators in the expression above. 
While they are indeed present, they do not make any difference to our discussion here and 
hence we do not explicitly show them. 
The masses of the $\hat{\Delta}_3$, $\hat{\Delta}_6$ and 
$\hat{\Delta}_{(3,2)}$ multiplets are obtained as follows:
\be
M_{\Delta_3}&=&m_{\Delta}-\frac{6 \lambda_{\Delta} v_{\Sigma}}{\sqrt{30}}
\,,\nn \\
M_{\Delta_6}&=&m_{\Delta}+\frac{4 \lambda_{\Delta} v_{\Sigma}}{\sqrt{30}}  
\,,\nn \\
M_{\Delta_{(3,2)}}&=&m_{\Delta}-\frac{\lambda_{\Delta} 
v_{\Sigma}}{\sqrt{30}}
\,,
\ee
where $\lambda_{\Delta}=\lambda_{\Delta}'+\lambda_{\Delta}''$. Here again we see 
that we can tune one of the multiplets to be at TeV scale. The masses 
of other multiplets present in $\hat{15}_H$ are then of GUT scale.

\subsection{Constraints from Gauge Coupling Unification}

In the following we demand that the SM gauge coulings unify at 
the GUT scale. This constrains the particle masses of the model. We  
try to find solutions of the particle mass spectra of our model 
where seesaw mediating particle can be below the TeV scale.  We find that in this 
supersymmetric version of extended SU(5) GUT model,  
the only seesaw mediating particle that can be 
made light turns out to be $\hat{\Delta}_3 \equiv (1,3,1)$.

We discuss this possibility in a little more detail. 
The role of the different SM representations 
belonging to $\hat{24}_F$ in the running of the gauge couplings 
have been discussed in the previous 
section. The representations coming from $\hat{15}_H$ (and 
$\hat{\bar{15}}_H$), are 
$(1,3,1)\oplus (3,2,1/6) \oplus (6,1,-2/3) 
\equiv (\hat{\Delta}_3, \hat{\Delta}_{(3,2)}, \hat{\Delta}_{6})  $. 
Therefore, while $\hat{\Delta}_3$ and $\hat{\Delta}_{(3,2)}$
affect the running of $\alpha_2^{-1}$, $\hat{\Delta}_{(3,2)}$ 
and $\hat{\Delta}_{6}$ lower the slope of $\alpha_3^{-1}$. 
All the three supermulitplets are non-trivial under U(1)$_Y$ and hence 
affect the running of $\alpha_1^{-1}$.  
For $M_{\Delta_3} \sim 100$ GeV, the gauge couplings are found to unify 
if $\hat{\rho}_3$, $\hat{\rho}_8$ and $\hat{\Sigma}_8$ are allowed to be 
at an intermediate scale, while all other non-SM particles are at the GUT scale. 
The GUT scale is related to $M_{\Delta_3}$ and $M_{\rho_3}$ as 
\be
\log M_G = 21.36 + 0.06\log M_{\Delta_3} - 0.396\log M_{\rho_3}
\,.
\label{eq:case11}
\ee
The Eq. (\ref{eq:case11}) shows that the GUT scale actually has a 
very mild dependence on the mass of the triplet scalar. On the other hand a small 
change to the GUT scale makes a big change to the mass of $\hat{\Delta}_3$. 
If one chooses $M_{\Delta_3} \sim 100$ GeV, then a GUT scale of 
$M_G \sim 10^{16}$ GeV can be obtained if the $\hat{\rho}_3$ 
mass is constrained to be $M_{\rho_3} \sim 10^{13}$ GeV.  
We will discuss the implications for this 
in the next section when we discuss neutrino masses and related phenomenology.
Note also from Eq. (\ref{eq:case11}) 
that smaller values of $M_{\Delta_3}$ are associated with smaller 
values of $M_G$. 
%The current limit on $M_G$ is around $\sim 10^{15.5}$ 
%GeV while the next generation megaton scale water Cherenkov 
%detectors should push up this limit to $M_G \sim 10^{x}$ GeV. 
%Therefore, observation of the $\hat{\Delta}_3$ at LHC 
%will certainly open up the possibility of observing proton 
%decay in the megaton water detectors, lending a smoking gun 
%signal for this model. 
 
The product of the masses of $\hat{\rho}_8$ and $\hat{\Sigma}_8$ are related to $\hat{\Delta}_3$ and 
$\hat{\rho}_3$ masses as 
\be
\log (M_{\rho_8} M_{\Sigma_8}) = -8.03 + 1.23\log M_{\Delta_3} + 1.3\log M^\rho_3
\,.
\label{eq:case12}
\ee
For $M_{\Delta_3} \sim 100$ GeV and $\hat{\rho}_3 \sim 10^{13.3}$ GeV,  the product of the 
$\hat{\rho}_8$ and $\hat{\Sigma}_8$ masses turns out to be $10^{11.7}$ GeV. 
Therefore in this framework, it is possible to have two 
particles, the seesaw mediating (1,3,1), and a SU(3) octet
(8,1,0) at the LHC scale. 
In fact, Eq. (\ref{eq:case12}) leads to the following situations:
\begin{itemize}
\item $\hat{\rho}_8$ is within the reach of the LHC and $\hat{\Sigma}_8$ has mass $\sim 10^{10}$ GeV,
\item $\hat{\Sigma}_8$ is within the reach of the LHC and $\hat{\rho}_8$ has mass $\sim 10^{10}$ GeV,
\item both $\hat{\rho}_8$ and $\hat{\Sigma}_8$ have masses in the intermediate regime.
\end{itemize}
In all the three possible cases mentioned above, the masses of $\hat{\Sigma}_8$ 
and $\hat{\rho_8}$ must  
below the GUT scale for unification of the gauge couplings. This requires that the 
$\hat{\Sigma_8}$ be split from the masses of $\hat{\Sigma}_0$ and $\hat{\Sigma}_3$, 
which are close to the GUT scale. Similarly, we require $M_{\rho_3}$ and $M_{\rho_8}$ 
masses to be split from  the mass of $\hat{\rho}_{(3,2)}$ which is constrained to be 
close to the GUT scale. The 
$\hat{\rho}_{0}$ mass is unconstrained from gauge coupling 
contraints. We have discussed the mass spectra of $\hat{24}_H$ and 
$\hat{24}_F$ in the previous subsection and have shown that such a 
splitting can be consistently obtained from the superpotential.

%%%%%%%%%%%%%%%%%%%%%%%%%%%%%%%%
\begin{figure}[t]
\begin{center}
\includegraphics[width=0.5\textwidth,angle=270]{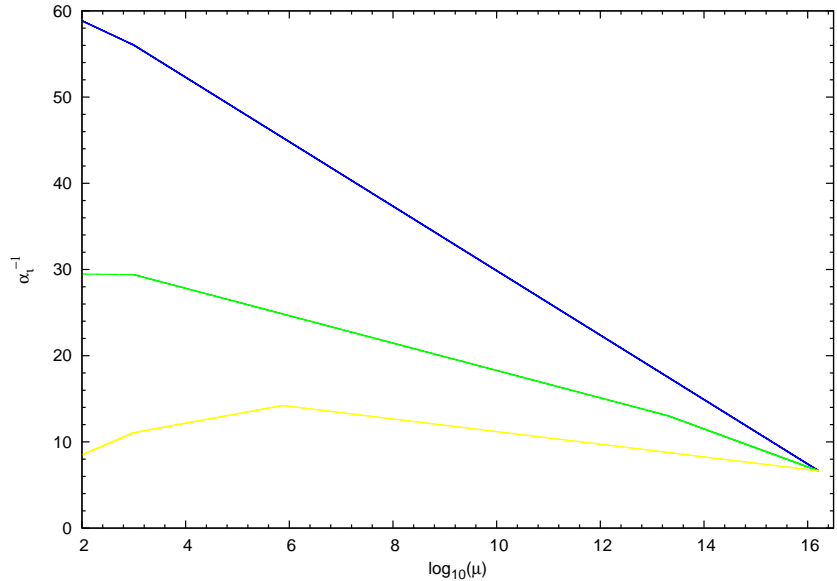}
\end{center}
\caption{\label{fig:rgsadssu5}
Gauge coupling running for the symmetric adjoint SUSY SU(5) with 
$M_{SUSY}=1$ TeV, $M_{\Delta_3}=100$ GeV, $M_{\rho_3}=10^{13.3}$ GeV, and 
$M_{\rho_8}=M_{\Sigma_8}=7.7\times 10^5$ GeV. 
}
\end{figure}
%%%%%%%%%%%%%%%%%%%%%%%%

%%%%%%%%%%%%%%%%%%%%%%%%%%%%%%%%
\begin{figure}[t]
%\begin{center}
\includegraphics[width=0.5\textwidth,angle=0]{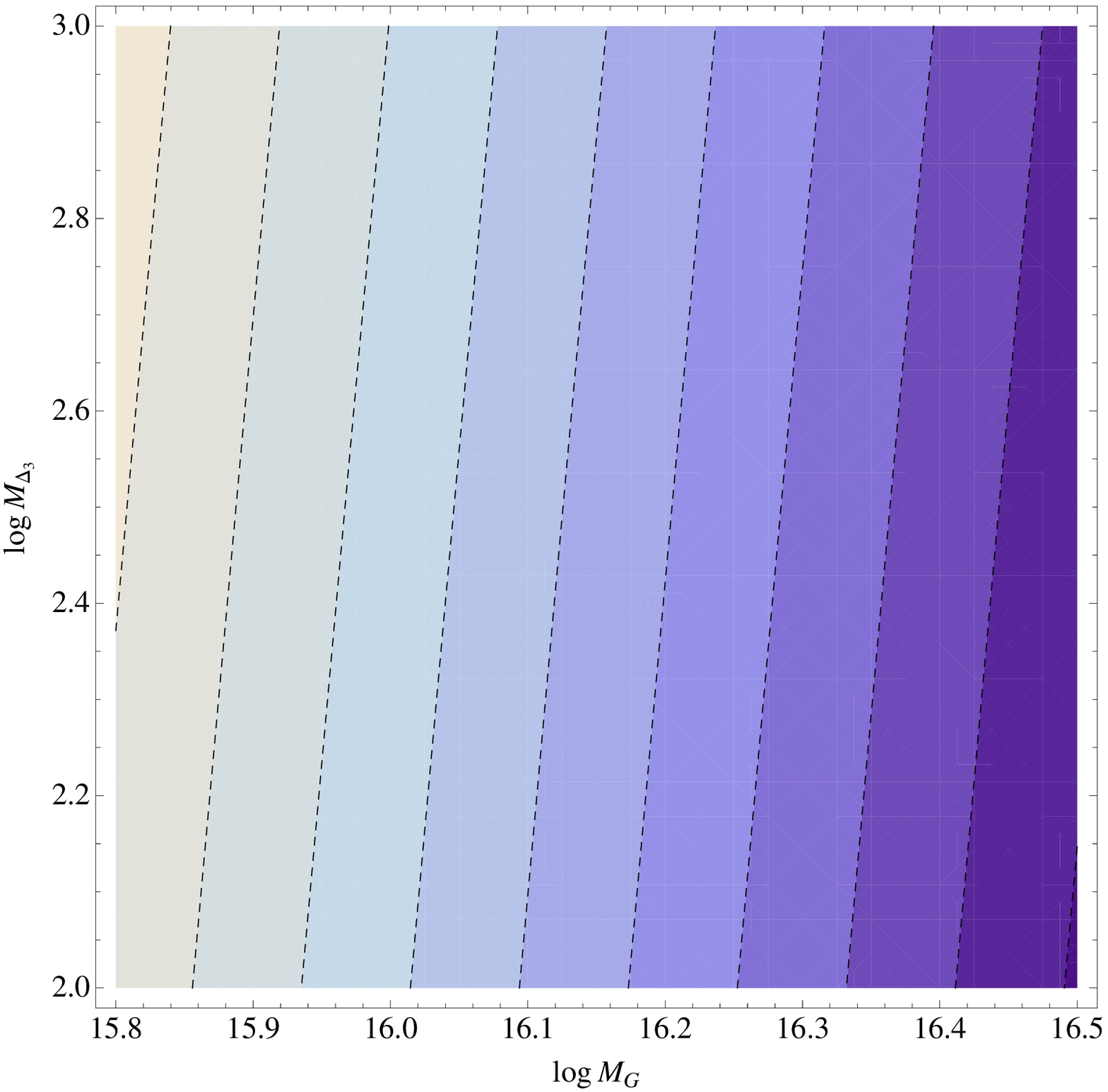}
\includegraphics[width=0.5\textwidth,angle=0]{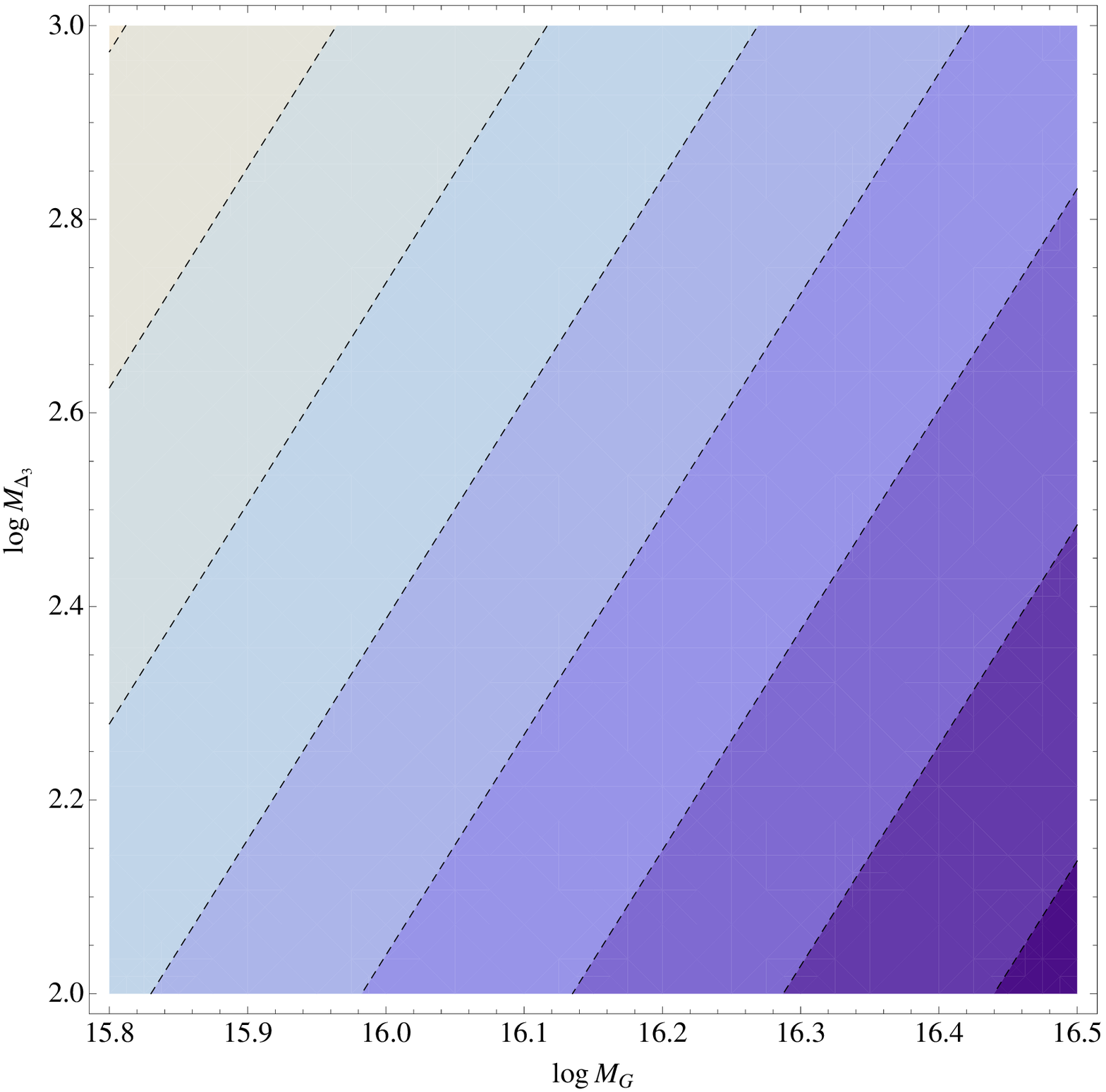}
%\end{center}
\caption{\label{fig:cont}
Left panel shows contours of 
$M_{\rho_3}$ for which one gets unification in the 
$\log M_G$ -$\log M_{\Delta_3}$
plane.  The leftmost dashed line corresponds to $M_{\rho_3}=10^{14.4}$ GeV, 
while the rightmost dashed line is for $M_{\rho_3}=10^{12.6}$ GeV. The value 
of $\log M_{\rho_3}$ increases by unit of 0.2 for every dashed line. 
The right panel of the figure shows the contours of 
$(M_{\rho_8} M_{\Sigma_8})$ in the $\log M_G$-$\log M_{\Delta_3}$ 
plane.  The right-most dashed line bordering the darkest part of the plot 
corresponds to $(M_{\rho_8} M_{\Sigma_8}) = 10^{11}$ GeV and each for 
each subsequent dashed line $\log(M_{\rho_8} M_{\Sigma_8})$ 
increases by 0.5.  
}
\end{figure}

%%%%%%%%%%%%%%%%%%%%%%%%

%%%%%%%%%%%%%%%%%%%%%%%%%%%%%%%%
%%%%%%%%%%%%%%%%%%%%%%%%

We show in Fig. \ref{fig:rgsadssu5} the running of the gauge couplings for the 
symmetric adjoint SUSY SU(5) model where we choose 
$M_{SUSY} = 1$ TeV, $M_{\Delta_3} = 100$ GeV, $M_{\rho_8} = M_{\Sigma_8}= 7.7 \times  10^{5}$ GeV, 
$M_{\rho_3}=10^{13.3}$ GeV and all other particles at the GUT scale. The GUT scale 
in this framework turns out to be $M_G=10^{16.2}$ GeV, 
being mainly determined by the 
mass of $\hat{\rho}_3$ (cf. Eq. (\ref{eq:case11})) along with a very weak dependence on 
the mass of $\hat{\Delta}_3$. The coupling constant at the GUT scale turns out to be 
$\alpha_G^{-1} = 6.7$, which is considerably lower than what one usually gets 
in minimal SUSY SU(5).

In the left panel of Fig. \ref{fig:cont} we show the contours for the value of 
$M_{\rho_3}$ for which one gets unification in the $\log M_G$ -$\log M_{\Delta_3}$
plane.  The leftmost dashed line corresponds to $M_{\rho_3}=10^{14.4}$ GeV, 
while the rightmost dashed line is for $M_{\rho_3}=10^{12.6}$ GeV. The value 
of $\log M_{\rho_3}$ increases by unit of 0.2 for every dashed line, 
as we move upward from 14.4 which borders the lightest color of the figure to 
12.6 which borders the darkest one (almost at the lower-right edge of the plot). This 
covers the range of $M_G$ from $10^{15.8}$  
to $10^{16.5}$ GeV. 
The plot shows that for $M_{\rho_3} \ltap  10^{13.2}$ GeV, $M_G \gtap 10^{16.2}$ GeV, 
while all values of $M_{\Delta_3}$ in the LHC range are possible. 
The right panel of Fig. \ref{fig:cont} shows the contours of 
$(M_{\rho_8} M_{\Sigma_8})$ in the 
$\log M_G$ -$\log M_{\Delta_3}$
plane.  
The right-most dashed line bordering the darkest part of the plot 
corresponds to $(M_{\rho_8} M_{\Sigma_8}) = 10^{11}$ GeV and for 
each subsequent dashed line $\log(M_{\rho_8} M_{\Sigma_8})$ 
increases by 0.5.  
Note that Eq. (\ref{eq:case12}) shows that fixing $(M_{\rho_8}  M_{\Sigma_8})$ 
fixes the value of $M_{\rho_3}$ for a given value of $M_{\Delta_3}$, and therefore 
both panels in this figure are equivalent way of showing the parameter region.

\section{Phenomenological Consequences of the Model}

\subsection{Neutrino Masses and Mixing}

In this subsection we discuss neutrino mass generation in our symmetric adjoint SUSY SU(5) 
model. The part of the superpotential  
which is involved in the generation of neutrino mass is
\be
{\cal W}_\nu  = {\cal W}_\nu^Y + {\cal W}_\rho^M + {\cal W}_\Delta
\,,
\label{eq:wnu}
\ee
where ${\cal W}_\rho^M$ is given in Eq. ({\ref{eq:wrhom}) and ${\cal W}_\nu^Y$ is the 
Yukawa part of the superpotential given by
\be
{\cal W}_\nu^Y=
{Y_\Delta}_{\alpha\beta}\hat{\overline{5}}_{F_\alpha}\hat{\overline{5}}_{F_\beta}\hat{15}_H
+Y_\alpha \hat{\overline 5}_{F_\alpha}\hat{24}_F\hat{5}_H
+ \frac{1}{\Lambda}\hat{\overline{5}}_F \bigg(Y_\alpha^1 \hat{24}_F \hat{24}_H +
Y_\alpha^2 \hat{24}_H \hat{24}_F +
Y_\alpha^3 Tr(\hat{24}_F \hat{24}_H) \bigg )  \hat{5}_H
%+m_{\rho} Tr[24_F]^2
%+ \lambda_\rho  Tr[24_F^2 24_H]
\,,
\label{eq:sup}
\ee
where we have kept the higher dimensional terms. The first term gives 
rise to the type II seesaw Majorana mass term when $\Delta_3$ in $\hat{15}_H$ 
acquires a VEV $v_\Delta$. The other terms in Eq. (\ref{eq:sup}) give 
contribution to the type I and type III seesaw due to 
electroweak symmetry breaking. 
The part ${\cal W}_\Delta$ in Eq. (\ref{eq:wnu}) contains the Higgs part relevant for type-II 
seesaw and is given by
\be
{\cal W}_\Delta = {\cal W}_\Delta^M + 
\mu_{\Delta}\hat{\overline{5}}_H\hat{15}_H\hat{\overline{5}}_H^T 
+ \mu_{\bar \Delta}\hat{{5}}_H^T\hat{\overline{15}}_H\hat{{5}}_H
\,,
\label{eq:type2higgs}
\ee
where we have kept only the renormalizable terms. The first term in Eq. (\ref{eq:sup}) gives a 
contribution $m_{II}=Y_\Delta v_\Delta$ to the low energy neutrino mass matrix. 
One can use the Eq. (\ref{eq:higgstrip}) and Eq. (\ref{eq:type2higgs}) 
to get the triplet Higgs VEV $v_\Delta$ as
\be
v_\Delta =  \frac{Y_\Delta \mu v_u^2}{M_{\Delta_3}^2}
\,,
\ee
where
\be
\mu = \mu_{\overline{\Delta}} M_{\Delta_3}
\,.
\ee
Hence, in this model the type II seesaw contribution to neutrino masses comes out to be 
\be
m_{II} = \frac{Y_\Delta \mu v_u^2}{M_{\Delta_3}^2}
\,
\label{eq:m2}
\ee
where $v_u$ is the VEV of the standard Higgs $H_u$. Note that 
the coupling $\mu_{\Delta}$ is dimensionless. 
There are of course several higher dimensional terms such 
as 
\be
{\cal W}_\Delta^{NR} &= &
\frac{\mu_{\Delta}^1}{\Lambda}\hat{\overline{5}}_H\hat{15}_H\hat{\overline{5}}_H^T\hat{24}_H 
+ \frac{{Y_\Delta}_{\alpha\beta}}{\Lambda}\hat{\overline{5}}_{F_\alpha}\hat{15}_H\hat{\overline{5}}_{F_\alpha}^T
\hat{24}_H 
+ \frac{1}{\Lambda}\hat{15}_H\hat{\overline 15}_H Tr(\hat{24}_H^2) 
+ \frac{1}{\Lambda}\hat{\overline 5}_{F_\alpha} \hat{15}_H \hat{\overline 5}_H \hat{24}_F 
\nn\\+ &&\frac{1}{\Lambda}\hat{15}_H \hat{\overline 15}_H Tr(\hat{24}^2_H)
+ \frac{1}{\Lambda}\hat{5}_H \hat{\overline 5}_H Tr(\hat{24}^2_H)
+ \frac{1}{\Lambda}\hat{5}_H \hat{\overline 5}_H \hat{15}_H \hat{\overline 15}_H
+ \frac{1}{\Lambda}\hat{5}_H \hat{\overline 5}_H \hat{24}^2_H {\overline 15}_H.....
\,,
\ee
which should be included in Eqs. (\ref{eq:sup}) and (\ref{eq:type2higgs}) for consistency. All these 
terms are suppressed by either $v_\Sigma/\Lambda$ (or $v_\Sigma^2/\Lambda)$
or $v_\Delta/\Lambda$ (or $v_\Delta^2/\Lambda$). Since 
$v_\Delta$ is constrained to be extremely small by the smallness of neutrino mass (cf. Eq. (\ref{eq:m2})), 
terms proportional to $v_\Delta/\Lambda$ (or $v_\Delta^2/\Lambda$) can be safely neglected. 
We have explicitly checked that the effect of all other terms can also 
be neglected since $v_\Sigma/\Lambda \sim 10^{-3}$ and hence they all give 
very small correction compared to the leading term, which anyway are present in 
the renormalizable part of the superpotential. 

As discussed before, in this model
the neutrino mass matrix can get contribution from type I and type III seesaw 
as well. Thus the complete neutrino mass matrix is given by 
\be
M_{\nu}= \frac{1}{2}m_{D_0}M_{\rho_0}^{-1}m_{D_0}^T + \frac{2Y_\Delta 
\mu v_u^2}{M_{\Delta_3}^2} + \frac{1}{2} m_{D_3}M_{\rho_3}^{-1}m_{D_3}^T 
\label{eq:mnu}
\ee
where $m_{D_0}$ and $m_{D_3}$ are as follows,
\be
m_{D_0}&=&-\frac{3v_uY}{\sqrt{30}}    + \frac{v_\Sigma v_u}{\Lambda} \bigg(
Y^3 + \frac{9}{30}(Y^1 + Y^2)\bigg)
\\
m_{D_3}&=&-\frac{v_uY}{\sqrt{2}} + \frac{3 v_\Sigma v_u}{\sqrt{60} \Lambda} \bigg(
(Y^1 + Y^2)\bigg)
\,,
\ee
and $M_{\rho_0}$ and $M_{\rho_3}$ given in Eq. (\ref{eq:rhomass})
are the masses of the singlet and triplet  
fermions of $\hat{24}_F$, respectively. 
The first, second and third terms in Eq. (\ref{eq:mnu}) are the contributions from 
type I, II and III seesaw, respectively. 
Since $M_{\Delta_3} \sim 100$ GeV in this model, 
 $Y_\Delta \mu \sim M_\nu$. Therefore, for $Y_\Delta \sim 1$ and 
 $\mu \sim 0.1$ eV, we get significant contribution to $M_\nu$ from type II seesaw. 
We have seen that gauge coupling unification 
gives $M_{\rho_3} \sim 10^{13}$ GeV. Therefore, 
the contribution to the neutrino mass 
matrix from type III seesaw is also expected to be significant 
under the natural assumption of $Y\sim 1$. Gauge coupling unification 
puts absolutely no constraint on the mass of $\rho_0$. We have seen 
in Eq. (\ref{eq:rhomass}) that the mass of $\rho_0$ can be 
tuned to any desired value. Therefore, we have type I plus 
type II plus type III seesaw in this model.

\subsection{Collider Signatures and Lepton Flavor Violation}

In the previous section we have seen that the $\hat{\Delta}_3$ belonging 
to the $\hat{15}_H$ representation is predicted to be of 100 GeV mass range
in our model. This opens up the possibility of testing the neutrino mass generation 
mechanism  at LHC by directly observing the type II seesaw mediating $\Delta_3$. 
The potential of testing type II seesaw at LHC has been extensively 
studied in the literature \cite{aguila, type2lhc, han}. The best way to probe type II 
seesaw is by observing the doubly charged Higgs scalar through its 
decay modes
\be
\Delta^{++} &\to& l^{+}l^{+} \nn \\ \nn
\Delta^{++} &\to& W^{+}W^{+} \\ \nn
\Delta^{++} &\to &\Delta^{+}W^{+} \\ \nn
\Delta^{++} &\to& \Delta^{+} \Delta^{+} \\ 
\Delta^{++} &\to& \tilde{\Delta}^+ \tilde{\Delta}^+ 
%\Delta^{++} \to \tilde{H}^{+} \tilde{H}^{+} \\ \nn
%\Delta^{++} \to \tilde{H}/\tilde{h} \tilde{H}^+ \\ \nn
%\Delta^{++} \to \tilde{\Delta}^+ \tilde{\lambda}^+
\,,
\label{eq:dspp}
\ee
where $\Delta^{++}$ and $\Delta^+$ are respectively the doubly and singly 
charged Higgs scalar in the mass basis. This model has two Higgs doublets and 
two Higgs triplets at the electroweak scale. Therefore, we have 2 doubly 
charged Higgs, 3 singly charged Higgs and 7 chargeless Higgs - 4 of which 
are CP even and 3 which are CP odd.  The branching ratios of the various 
competing channels in Eq. (\ref{eq:dspp}) depends crucially on the 
VEV $v_\Delta$, mass $M_{\Delta_3}$ and Yukawa coupling $Y_\Delta$. 
For $M_{\Delta_3} \simeq 300$ GeV, the decay of $\Delta^{++}$ into the 
dilepton channel dominates over $W$ production channels if  
$v_\Delta \ltap 10^{-4}$ GeV \cite{type2lhc, han}. This limiting value of $v_\Delta$ 
increases as $M_{\Delta_3}$ increases.  The decay width of the dilepton 
channel depends on the strength of the Yukawa coupling matrix $Y_\Delta$.  
More importantly, the lepton flavor in the final state of the dilepton channel 
depends on the individual matrix elements of $Y_\Delta$. This 
provides a one-to-one correspondence between the neutrino mass matrix 
and LHC signatures in pure type II seesaw models. However, in our 
model neutrino masses could have contributions from all three types of seesaw. 
Therefore, it is possible that we do not have a one-to-one correspondence between 
the neutrino experiments and the signal at LHC. One can turn this argument around and say 
that any discrepancy between the neutrino and LHC experiments would point 
towards a hybrid seesaw model with scalar triplets, such as ours. 

The particles $\tilde{\Delta}^+$ in Eq. (\ref{eq:dspp}) are charginos
in the mass basis. Including the contribution coming from the Higgs triplet field and the MSSM charginos, in
our model we have 
3 singly charged charginos ($\tilde{\Delta}^+$). In addition we also have  one doubly charged chargino ($\tilde{\Delta}^{++}$ ) and 6 neutralinos ($\tilde{\Delta}^0$) in our model. 
%which in our model 
%are a mixed state comprising of $\tilde{H}_1^+$, $\tilde{\bar{H}}_1^+$, 
%$\tilde{\lambda}^+$, ${\tilde{\lambda}^-}C$ and $\tilde{\Delta}^+$. 

The singly charged Higgs scalar decays via
\be
\Delta^{+} &\to&  l^{+}\nu \,,\nn \\
\Delta^{+} &\to&  \Delta^0W^+
\,.
\ee
Since we have imposed supersymmetry we have 100 GeV mass range 
fermionic partners of the charged Higgs particles. Some of their decay modes are 
\be
\tilde{\Delta}^{++} & \to &  \tilde{l}l\,, \nn \\ \nn 
\tilde{\Delta}^{+}  & \to  & \tilde{l}\nu/ l\tilde{\nu} \,,\\ \nn
\tilde{\Delta}^{++}  & \to &  W^{+}\tilde{\Delta}^+\,, \\ \nn
\tilde{\Delta}^{++}  & \to  & \tilde{\Delta}^+ \Delta^+ \,,\\ \nn
\tilde{\Delta}^0  & \to  & \tilde{\Delta}^0 \Delta^0 \,, \\ \nn 
\tilde{\Delta}^{+}  & \to  & \tilde{\Delta}^{+}  \Delta^0 \,,\\ 
\tilde{\Delta}^{+}  & \to  & \tilde{\Delta}^0  \Delta^{+}\,.
\ee
From Eq. (\ref{eq:spot}) one can see that the 
octet fermions $\rho_8$ and $\tilde{\Sigma}_8$ decay via channels such as 
\be
\rho_8 &\to& d^C T\,,\nn \\
\rho_8 &\to& q^C \bar{\Delta}_6\,,\nn \\
\tilde{\Sigma}_8 &\to& \tilde{\Delta}_{(3,2)} \overline{\Delta}_{(3,2)}
\ee
while the octet scalars decay through
\be
\tilde{\rho_8} &\to& \tilde{q}^C \overline{\Delta}_6\,,\nn \\
\tilde{\rho_8} &\to& {d}^C \tilde{T}/\tilde{d^C} T\,,\nn \\
{\Sigma}_8 &\to& \tilde{\Delta}_{(3,2)} \tilde{\overline{\Delta}}_{(3,2)}
\ee
Since all these modes involve a GUT scale particle, which have to be 
produced off-shell,  the decay of the 
octet fermions and scalar happen via effective operators which are suppressed 
by the GUT scale and hence these particles have very long lifetimes.

\subsection{Proton Decay}

There are no additional proton decay mediating diagrams in our model. If we assume 
that the contribution of dimension five operators in SUSY SU(5) is negligible 
\cite{bps02}, then proton decay is mediated only by dimension six 
operators involving particles with masses of the GUT scale, {\it viz.}, 
the superheavy gauge bosons $X$ and $Y$, the SU(3) triplets $T$ and $\overline{T}$. 
Therefore, the decay width for proton decay in our model is same as that in 
minimal SUSY SU(5) where contribution from dimension five operators are 
negligible. This has been discussed widely in the literature (for a review see \cite{pd}). 
The predicted decay width for the following channels  are \cite{perez}
\be
\tau (p \to \pi^+ \bar{\nu}) = 1.44\times 10^{-31}\bigg(\frac{(M_G/GeV)^4}{\alpha_G^2}\bigg)~{\rm years}
\,,
\\
\tau (p \to K^+ \bar{\nu}) = 4.31\times 10^{-30}\bigg(\frac{(M_G/GeV)^4}{\alpha_G^2}\bigg)~{\rm years}
\,.
\ee
%Since the GUT scale is predicted to be greater than $10^{16.3}$ GeV for 
%$M_{\rho_3}\ltap 10^{13.3}$ GeV in our model, the lifetime for the 
%following channels is predicted to be
%\be
%\tau (p \to \pi^+ \bar{\nu})\gtap 4.0\times 10^{35} ~{\rm years}
%\,
%\\
%\tau (p \to k^+ \bar{\nu})\gtap 1.2\times 10^{37} ~{\rm years}
%\,.
%\ee
%This should be testable in large scale future detectors 
%such as megaton water Cherenkov detectors or large Liquid Argon Projection 
%Chambers 
%\cite{megaton}.  
The relation between $M_{\rho_3}$, $M_G$ and $M_{\Delta_3}$ 
has been discussed in the previous section. In particular, one can see this in 
Eq. (\ref{eq:case11}) and Fig. \ref{fig:cont}.  Since $\hat{\Delta}_3$ is testable 
at LHC and both $\Delta_3$ and $\rho_3$ are related to neutrino masses, it 
could be possible to related the predictions at proton decay signatures at 
large scale future detectors \cite{laguna} 
with the results from the LHC and 
neutrino oscillation experiments.

\section{Conclusions}

The minimal SU(5) cannot explain the presence of neutrino masses and 
must be necessarily extended. Addition of SU(5) singlets 
gives rise to type I seesaw, 
addition of $15_H$ 
of SU(5) gives type II seesaw masses, while the 
extension with $24_F$ 
makes it possible to generate type III seesaw mass 
term. With the LHC running, it is pertinent to expect that one could probe 
seesaw at this collider experiment. An obvious pre-requisite for this is that 
the mass of the seesaw mediating particle should be within the reach of the LHC, 
and hence should have masses within 1 TeV.  In this context it was rather 
exciting to note that in the SU(5) model extended with the $24_F$ multiplet, 
known as adjoint SU(5) in the literature, gauge unification 
imposed that the mass of the seesaw mediating $\rho_3$ should be close to 
100 GeV range, making it accessible at LHC. However, 
once this model is supersymmetrized, we found that mass of $\rho_3$ 
should be close to the GUT scale in order to get gauge coupling unification.  
Therefore, seesaw in the SUSY adjoint SU(5) cannot be probed at the LHC. 
The SU(5) GUT model extended by a $15_H$ also does not predict 
100 GeV scale type II seesaw mediating triplet scalar masses. 

We proposed a SUSY SU(5) GUT model extended with 
a $\hat{24}_F$ matter chiral field, and $\hat{15}_H$ and $\hat{\bar{15}}_H$ 
Higgs chiral fields. We call this model the {\it symmetric adjoint SUSY SU(5)}. 
In principle, this model can get contribution from type I, II 
as well as III seesaw mechanisms. We showed that it is possible to consistently 
predict masses for triplet scalar and triplet fermion in the 100 GeV and 
$10^{13}$ GeV range respectively, while the mass of $\rho_0$ is unconstrained. 
This gives type I plus type II plus type III seesaw 
mass term for the neutrinos.  The triplet fermion masses in the $10^{13}$ GeV 
range allows for type III seesaw with Yukawa couplings of the order of 1. The 
triplet scalars with masses $\sim 100$ GeV range can be produced at the LHC. 
We briefly discussed the collider phenomenology of this model and 
prediction for proton decay.

\end{document}